\documentstyle[preprint,aps]{revtex}
\date{\today}
\begin{document}
\draft
\title{\bf
The Energy-Energy Correlation Function of the Random Bond
Ising Model in Two Dimensions}
\author{K. Ziegler}
\address{Institut f\"ur Theorie der Kondensierten Materie,
Universit\"at Karlsruhe, Physikhochhaus, D-76128 Karlsruhe, Germany}
\maketitle
\begin{abstract}
The energy-energy correlation function of the two-dimensional Ising
model with weakly fluctuating random bonds is evaluated in the large scale
limit. Two correlation lengths exist in contrast to one correlation length
in the pure 2D Ising model: one is finite whereas the other is divergent at
the critical points. The corresponding exponent of the divergent correlation
length is $\nu_e=1/2$ in contrast to the pure system where $\nu_e=1$.
The calculation is based on a previously developed effective field theory
for the energy density fluctuations.
\end{abstract}
The two-dimensional Ising model\cite{1} with random bonds is an interesting
example for the competition of thermal fluctuations and quenched
disorder. According to the Harris criterion\cite{2} neither of these two
effects dominates over the other near the critical
point. This picture is supported by a number of investigations for
weak disorder, treated in perturbation theory with respect to the
variance of the bond distribution, which came to the conclusion that
the thermodynamic properties are not dramatically changed by quenched
disorder\cite{3,4,5,6}. This is a consequence of the fact that the fixed
point of the pure Ising model is stable against disorder (at least in one-loop
order). The effect of disorder on the power law of the spin-spin correlation
$\langle S_RS_0\rangle\sim R^{-1/4}$ of the pure system is controversal. While
some authors claim there is no effect on the exponent $1/4$
\cite{4,5,6} others find a change for the correlation function to
$\exp\{-const.[\log(\log R)]^2\}$\cite{3}. On a naive basis one expects a
reduction of the correlation near the critical point due to the additional
fluctuations of the random bonds. Numerical simulations\cite{7,8}
do not indicate a significant change of the correlation function.

Apart from the perturbative approaches it was found more recently in a
non-perturbative treatment that disorder modifies the phase diagram
of the two-dimensional Ising model\cite{9,10}. In particular, disorder
creates a new phase between the ferro- and the paramagnetic phase. However,
the width of this phase is small and vanishes for any finite order in
perturbation theory. For the same reason it is rather unlikely that this
this phase can be observe in a numerical simulation.
Nevertheless, there is a rigorous proof for its
existence\cite{11}. This phase is characterized by a new order parameter.
It was not possible in previous studies to understand its role in terms of
the thermodynamic functions. However, it was related to the inverse
correlation length of the average spin-spin correlation function by
Braak\cite{12}. In the following it will be shown that it also plays the
role of an inverse correlation length of the average energy-energy correlation
function.

The Ising model describes a system of discrete spins $S_r=\pm 1$ on a
lattice. In this article a square lattice will be considered.
The spins on nearest neighbor sites of the lattice are coupled with the
energy
\begin{equation}
{\cal H} =-\sum_{r,r'}J_{r,r'}S_rS_{r'}
\end{equation}
where $J_{r,r'}$ is a random coupling variable. A statistical system of
spins at the inverse temperature $\beta$ is defined by the partition
function
\begin{equation}
Z=\sum_{\{S_r\}}\exp(-\beta{\cal H} ).
\end{equation}
Using this quantity the energy-energy correlation function (EECF) reads
\begin{equation}
\langle{\cal E}_r{\cal E}_{r'}\rangle={1\over\beta^2}\sum_{r'',r'''}
J_{r,r''}J_{r',r'''}{\partial\over\partial J_{r,r''}}
{\partial\over\partial J_{r',r'''}}\log Z.\label{ec}
\end{equation}
In order to calculate $Z$ a number of mappings and approximations is useful.
Firstly, the statistics of Ising spins can be mapped onto the statistics
of closed polygons which excludes multiple occupation of links between
lattice
sites. This leads to $Z=Z_0Z_P$ where $Z_0$ is a spin-independent part
and $Z_P$ the partition function of the closed polygons. Secondly, $Z_P$
is equivalent with the partition function of non-interacting fermions with
four colors. As a consequence, $Z_P^2$ is as fermion determinant.
An approximation for weak disorder and large scales reads as a fermion
determinant for complex fermions with two colors
$Z_P^2\approx\det(H_0+\delta y\sigma_3)$.
The Fourier components of the lattice matrix $H_0$ are
\begin{equation}
{\tilde H}_0=m\sigma_3+{\bar y}_c(\sigma_2\sin k_1+\sigma_1\sin k_2).
\end{equation}
$\delta y_r$ represents the random fluctuations of the couplings with
variance $g$ and
$m=y-y_c$, $y=\tanh(\beta J)$,  $y_c=\sqrt{2}-1$. $\sigma_j$
are the Pauli matrices.
Introducing the Green's function $G(\varepsilon)=(H_0+\delta y\sigma_3+
i\varepsilon\sigma_0)^{-1}$ the EECF reads in this approximation
\begin{equation}
\langle{\cal E}_r{\cal E}_{r'}\rangle
\approx\lim_{\varepsilon\to0}
\langle({\bar y}+\delta y_r)({\bar y}+\delta y_{r'})Tr [G_{r,r'}G_{r'
,r}]\rangle\label{ecg}
\end{equation}
with a positive constant ${\bar y}$.
Naively, one would neglect in a weak disorder approximation the $\delta y$
contributions in front of the $G$'s. However, we will see in the following
that these terms are important. This is the first step for the derivation
of the EECF in the disordered Ising system.

As given in (\ref{ec}) the fluctuations of the random coupling $J_{r,r'}$ are
conjugate to the energy density ${\cal E}_r$. Therefore, it is convenient
to transform the distribution of the fluctuations $\delta y$ into a
distribution of the energy fluctuations. The latter was performed in
Ref. \cite{9} by mapping the random bond fluctuations to a random
matrix
\begin{equation}
\delta y_r\sigma_3\to Q_r.
\end{equation}
$Q_r$ is a Hermitean $2\times2$ matrix with random matrix elements.
It describes the energy density fluctuations. For instance, the average
energy density is proportional to the average trace of $Q$.
Without discussing the structure of this random matrix model (the details
can be found in\cite{9}) we turn directly to the saddle point approximation of
the model. It is sufficient to consider the upper left $2\times2$ block
matrix $Q_r=Q_0+\delta Q_r$ where $Q_0$ is given by a saddle point
approximation\cite{9} with
\begin{equation}
Q_0 =-{i\over 2}\eta\sigma_3-{1\over 2}m_s\sigma_0.
\end{equation}
$\eta$ and $m_s$ are solutions of the saddle point equations
\begin{equation}
\eta =\eta g I
\end{equation}
\begin{equation}
{\bar m}=m+m_s={m\over 1 +gI}
\end{equation}
with the integral
$I=2\int\lbrack {\bar m}^2 +\eta^2+\vert\kappa\vert^2\rbrack^{-1}d^2k
/(2\pi)^2$. The solution, which describes the new intermediate phase,
is given by $\eta\ne0$ as
\begin{equation}
\eta^2=e^{-2\pi{\bar y}_c^2/g}-m^2/4
\end{equation}
and ${\bar m}=m/2$. $\eta$ vanishes at the critical points $m_c=\pm
2e^{-\pi{\bar y}_c^2/g}$ and is zero for $m^2>m_c^2$.

The EECF can be expressed by the Gaussian fluctuations around the saddle
point solution $Q_0$ as
\begin{equation}
\langle{\cal E}_r{\cal E}_{r'}\rangle\approx
{1\over {\cal N}}\int[{\bar y}Tr(Q_r)+Tr(Q_r^2)][{\bar y}Tr(Q_{r'})+
Tr(Q_{r'}^2)]\exp[-(\delta Q,{\hat I}\delta Q)]\prod d\delta Q
\label{ex}\end{equation}
with normalization ${\cal N}$.
The first order terms in Q are the contribution from ${\bar y}G$ whereas
the quadratic terms $Tr(Q^2)$ are a result of $\delta yG$ in (\ref{ecg}).
Higher order terms in the exponential function are neglected in this
approximation. Taking them into account (e.g. third order terms in $Q$)
could lead to a renormalization of ${\hat I}$. On the other hand, the
saddle point approximation corresponds with the $N\to\infty$ limit of a
more general model\cite{10}. Fluctuations appear as corrections with
$1/N^{-1/2}$ for each $\delta Q$. Therefore, expression (\ref{ex}) should
be a good approximation in terms of the $1/N$-expansion.

The eigenvalues $\lambda_1$,..., $\lambda_3$ of the $4\times4$ matrix
${\hat I}(k)$ are positive whereas $\lambda_4$ vanishes like $\eta^2$ if
one approach the critical points $m_c$:
\begin{equation}
\lambda_4\sim const.|m + m_c|+b k^2+o(k^3).
\end{equation}
The mode of $\lambda_4$ is $\delta Q_{11}-\delta Q_{22}$. Therefore, the
terms with $Tr(Q)=m_s+\delta Q_{11} +\delta Q_{22}$ are {\it not} critical
in contrast to terms with $Tr(Q^2)$. This fact explains why the $\delta y$
contributions are important in
the expression of the EECF in (\ref{ecg}). The mode $\delta Q_{11}+\delta
Q_{22}$, which corresponds to $\lambda_3$, has a characteristic finite
correlation length $\xi_3\sim const.\lambda_3^{-1/2}$.
Furthermore, there is a divergent correlation length from the
correlation of $Tr(Q_r^2)$ which is proportional to $|\eta|^{-1}$, the
order parameter of the intermediate phase. It reads in terms of $m$
$\xi_4\sim\xi_0|m + m_c|^{-1/2}$.

Thus, the EECF is characterized by two different length scales. One is the
{\it finite} correlation length $\xi_3$, the other is the {\it divergent}
correlation length $\xi_4$ which is related to the order parameter
$|\eta|^{-1}$ of the new phase. In contrast to this, the pure system has
only one correlation length which diverges at the critical point $m=0$ as
$\xi_{pure}\sim |m|^{-1}$.

In conclusion, quenched disorder reduces significantly the correlation length
of the energy fluctuations in the two-dimensional Ising model (see table 1).
This result indicates that disorder is not only marginal in this model.
A similar reduction is expected for the spin-spin correlation function in
opposition to the controversal results obtained previously in
perturbation theory\cite{3,4,5,6}. However, Braak\cite{12} found in saddle
point approximation that the
correlation length of the average spin-spin correlation function
in the intermediate phase is proportional to $\eta^{-2}$; i.e., the
critical exponent for this correlation length is $\nu=1$ as in the pure
Ising system.

\begin{table}
\caption{Table of critical points and exponents for correlation lengths of
the energy-energy correlation function (with the constant $c=\pi{\bar
y}_c^2\approx0.54$).}
\begin{tabular}{llcdcc}
&         & pure   & disordered      \\
\tableline
& $m_c$   & 0      & $\pm 2e^{-c/g}$ \\
& $\nu_e$ & 1      &  1/2            \\
\end{tabular}
\end{table}

\begin{references}

\bibitem{1}
B.M. McCoy and T.T. Wu ``The Two-dimensional Ising Model''
Harvard University Press, Cambridge, Mass., 1973

\bibitem{2}
A.B. Harris, J.Phys. C7, 1671 (1974)

\bibitem{3}
V. Dotsenko and Vl.S. Dotsenko Adv.Phys. 32 (1983) 129

\bibitem{4}
B.N. Shalaev, Sov.Phys.Solid State 26, 1811 (1984)

\bibitem{5}
R. Shankar, Phys.Rev.Lett. 58, 2466 (1987)

\bibitem{6}
A.W.W. Ludwig, Nucl.Phys. B285, 97 (19987), Phys.Rev.Lett. 61, 2388
(1988)

\bibitem{7}
J.S. Wang, W. Selke, Vl.S. Dotsenko and V.B. Andreichenko, Physica A164
(2), 221-239 (1990)

\bibitem{8}
A.L. Talapov, V.B. Andreichenko, Vl.S. Dotsenko, L.N. Shchur,
in ``Computer simulations in condensed matter physics'' eds.
D.P.Landau, K.K.Mon, and H.B.Schuettler, (Springer, Heidelberg, 1991)

\bibitem{9}
K. Ziegler, Nucl.Phys. B344, 499 (1990)

\bibitem{10}
K. Ziegler, Europhys.Lett. 14, 415 (1991)

\bibitem{11}
K. Ziegler, Nucl.Phys. B285, 606 (1987)

\bibitem{12}
D. Braak, preprint Karlsruhe (cond-mat@babbage.sissa.it no. 9311045)
\end{references}
\end{document}